\documentclass[11pt]{article}
\usepackage{float}

\usepackage{amsmath}
\usepackage{amssymb, amsmath, amsthm}
\usepackage{mathrsfs}
\usepackage{tabulary}
\usepackage{rotating}
\usepackage[para]{threeparttable} 
\usepackage{enumerate}
\usepackage{setspace}
\usepackage{multirow}
\usepackage{paralist}
\usepackage{listings}
\usepackage[T1]{fontenc}
\usepackage[utf8]{inputenc}
\usepackage[english]{babel}
\usepackage{dsfont}
\usepackage{hyperref}
\usepackage{subcaption}
\usepackage{algorithm}
\usepackage{algpseudocode}
\usepackage{dsfont}
\usepackage{arydshln}

\DeclareMathOperator*{\argmax}{arg\,max}
\renewcommand{\d}{\mathsf{d}}
\newcommand{\m}{\mathsf{m}}
\newcommand{\g}{\mathsf{g}}
\newcommand{\E}{\mathsf{E}}
\renewcommand{\Pr}{\mathsf{P}}

\usepackage{graphicx}
\usepackage{url}
\usepackage{booktabs}
\usepackage{enumitem}
\newlist{subquestion}{enumerate}{1}
\setlist[subquestion,1]{label=(\alph*)}

\usepackage{authblk}
\usepackage{etoolbox}
\usepackage{longtable}
\usepackage[numbers,super,sort&compress]{natbib}
\usepackage[usenames,dvipsnames]{color}
\usepackage{natbib}
\usepackage{pdfpages}

\usepackage{caption}
\usepackage{hyperref}

\usepackage[hmargin=2.54cm,vmargin=2.54cm]{geometry}
\usepackage[utf8]{inputenc}

\newtheoremstyle{example}{}{}{}{}{\bfseries}{\smallskip}{\newline}{}
\theoremstyle{example}

\title{Learning Optimal Dynamic Treatment Regimes from Longitudinal Data}

\vspace{3cm}

\author[1]{Nicholas T. Williams, MPH}
\author[1]{Katherine L. Hoffman, MS}
\author[2]{Iv\'an D\'iaz, PhD}
\author[1]{Kara E. Rudolph, PhD}
\affil[1]{\footnotesize Department of Epidemiology, Mailman School of Public Health, Columbia University, New York, New York}
\affil[2]{\footnotesize Division of Biostatistics, Department of Population
  Health Sciences, New York University Grossman School of Medicine, New York, New York}
\author{}
\date{}

\begin{document} 

\maketitle

\begin{abstract}
Studies often report estimates of the average treatment effect (ATE). While the ATE summarizes the effect of a treatment on average, it does not provide any information about the effect of treatment within any individual. A treatment strategy that uses an individual’s information to tailor treatment to maximize benefit is known as an optimal dynamic treatment rule (ODTR). Treatment, however, is typically not limited to a single point in time; consequently, learning an optimal rule for a time-varying treatment may involve not just learning the extent to which the comparative treatments’ benefits vary across the characteristics of individuals, but also learning the extent to which the comparative treatments’ benefits vary as relevant circumstances evolve within an individual. The goal of this paper is to provide a tutorial for estimating ODTR from longitudinal observational and clinical trial data for applied researchers. We describe an approach that uses a doubly-robust unbiased transformation of the conditional average treatment effect. We then learn a time-varying ODTR for when to increase buprenorphine-naloxone (BUP-NX) dose to minimize return-to-regular-opioid-use among patients with opioid use disorder. Our analysis highlights the utility of ODTRs in the context of sequential decision making: the learned ODTR outperforms a clinically defined strategy.

\end{abstract}
\newpage
Epidemiological studies often report estimates of the average treatment effect (ATE), which is an estimand that summarizes the differences in an outcome of interest if all members of a population had been given a treatment or exposure compared to if all members of the population had been left untreated, unexposed, or treated in some other way. While the ATE can be a useful measure of whether a treatment is beneficial or harmful on average, it does not provide information about whether the treatment is beneficial or harmful for a particular individual. Indeed, it is possible that a treatment that is beneficial on average for a population, could benefit some individuals but harm others---in other words, exhibit treatment effect heterogeneity. 

Personalized medicine operates on the premise that treatment effect heterogeneity should be assumed, and aims to tailor clinical interventions to `the uniquely evolving health status of each patient', rather than adopting a one-size-fits-all approach.\cite{kosorok2021introduction,kosorok2019precision} The fundamental principle underlying personalized medicine---and, indeed, all clinical medicine---is that treatments should be administered to those who stand to benefit from them, while avoiding unnecessary treatment for those who would not benefit. A treatment strategy that deterministically assigns treatment as a function of an individual's history is known as a \textit{dynamic treatment rule}, a dynamic treatment rule that maximizes stratum-specific benefit to a patient is known as an \textit{optimal dynamic treatment rule} (ODTR).\cite{murphyODTR2003} For example, in a recent comparative effectiveness trial of two common treatments for opioid use disorder (OUD), an optimal treatment rule was learned in which patients who were homeless were found to have lower risk of return-to-regular-opioid-use (RROU) if treated with injection naltrexone, whereas patients with stable housing were found to have lower risk of relapse if treated with the alternative medication, buprenorphine-naloxone (BUP-NX).\cite{rudolph2021optimizing}

Treatment, however, is typically not limited to a single point in time. Rather, it may involve a series of decisions and adjustments. This is the case for treating opioid use with medication, where, for example, a clinician may choose to increase the prescribed dose over time. 
Consequently, learning an optimal rule for a time-varying treatment may involve not just learning the extent to which the comparative treatments' benefits vary across the clinical characteristics of individuals, but also learning the extent to which the comparative treatments' benefits vary as clinical characteristics or other relevant circumstances evolve within an individual. 

To our knowledge, much of the work on time-varying ODTRs has focused on the theory underlying identification of the causal parameters and estimation strategies.\citep{murphyODTR2003,zhangSequential2013,luedtkeSLodtr,DiazEnsembleTTE,zhaoBowl} Tutorials and/or software packages to facilitate implementation of these estimators for applied researchers are largely absent, which could, in part, explain why time-varying ODTRs have not been more widely adopted in epidemiology. In this paper we aim to address this gap by describing a doubly-robust approach (the DR-estimator or DR-learner) to estimate a time-varying ODTR from observational or randomized data. 

We begin by providing intuition for understanding static and dynamic treatment rules. We then explain how to use the DR-estimator to learn an ODTR for a single time-point, and then generalize the procedure to the time-varying setting. Last, we apply this framework to learn the ODTR for adapting weekly BUP-NX dose with respect to minimizing RROU over the first 6-weeks of treatment among patients with OUD. After learning the ODTR, we compare the estimated risk of RROU under the ODTR to the estimated risk of RROU using two previously defined BUP-NX dosing rules.\citep{rudolph2022dose}

\section*{Intuition}

We begin with some intuition for the single time-point setting. Denote $A$ as the treatment, and $Y$ the outcome on which we aim to assess the effect. Define $\d(V)$ as a function that takes an input of pre-treatment variables $V$ and maps it to a value of $A$. We call $\d(V)$ a \emph{treatment rule}. The possibly counterfactual outcome $Y_\d$ is then the hypothetical value of $Y$ if $A$ were assigned according to treatment rule $\d(V)$.

\subsection*{Dynamic treatment rules}

Imagine a hypothetical world where everyone receives the same value of a treatment or exposure. This type of intervention, or treatment rule, can be described as \emph{static}, because, in this alternate world, all members of the population receive the same treatment; regardless of the input, $\d$ always outputs the same value of $A$. In our application, one example of a static intervention would be to prescribe all patients the same dose of BUP-NX (scenarios 1 and 2 in Table \ref{tab:rule_examples}).

Now imagine another world where, instead of everyone receiving the same treatment, treatment is assigned to a subject as function of their measured history (denote this $H$); this intervention or rule is referred to as being \emph{dynamic}.\citep{hernanWhatIfDynamic} Consider, for example, the finding that individuals diagnosed with OUD and who are housing-unstable have lower risk of RROU when prescribed extended-release injection naltrexone versus when prescribed BUP-NX.\cite{rudolph2021optimizing} A relevant dynamic treatment rule may be one in which clinicians use their patient's housing status to decide which pharmaceutical treatment to prescribe. This dynamic rule could contain additional measured covariates, such as the patient's age or the length of time since their diagnosis with OUD. We may choose some or all of these variables in $H$, denoted as $V$, to be part of the dynamic rule. In other words, the output of $\d$ now changes according to the input of $V$ (scenario 3 in Table \ref{tab:rule_examples}). 

Treatment plans rarely can be reduced to a single decision; often, they vary over-time. For example, a physician treating OUD may be tasked with making a sequence of treatment decisions for increasing or not increasing BUP-NX dose for a single patient at every follow-up appointment. Consider a study with discrete time-points (i.e., days, weeks, months) where, at each time-point, a patient can either receive a dose increase or not. Instead of a single treatment rule $\d$, there is now a sequence of possible treatment rules for each time-point $\{1, ..., \tau\}: \mathbf{D} = [\d_1(V_1), \dots, \d_\tau(V_{\tau)}]$ where each rule $\d_t$ conditions on the observed history--including both baseline covariates and time-varying covariates--until the current time-point, $V_{t}$ (scenario 4 in Table \ref{tab:rule_examples}).

We note that defining a dynamic treatment rule as a rule that conditions on any variable in $H_{t}$ is only one possible definition. Others have defined dynamic treatment rules as treatment regimes that vary over time and condition only on time-varying covariates up until the current time-point; and defined static treatment rules as those that condition only on baseline information.\citep{cain2010start} Using this alternative definition, only scenario 4 in Table \ref{tab:rule_examples} would be considered a dynamic treatment rule.

\begin{table}[H]
    \centering\footnotesize
    \caption{Examples of static and dynamic treatment rules.}
    \begin{tabular}{>{\arraybackslash} m{120mm} c }
    \toprule
    Scenario & Treatment rule \\
    \midrule
    \begin{enumerate}
        \item All patients are prescribed BUP-NX at an initial visit. A dose of 16/4 mg/day is prescribed for all patients. BUP-NX dose is held constant throughout the course of treatment.
    \end{enumerate} & Static \\
    \midrule
    \begin{enumerate}
        \setcounter{enumi}{1}
        \item All patients are prescribed BUP-NX at an initial visit. All patients receive an initial dose 4/1 mg/day. The treatment plan includes a dose increase every two-weeks until a maximum dose of 24/6 mg/day is reached.
    \end{enumerate} & Static \\
    \midrule
    \begin{enumerate}
        \setcounter{enumi}{2}
        \item All patients are prescribed BUP-NX at an initial visit. Patients are either prescribed an initial dose of 4/1 mg/day or 16/4 mg/day depending on the level of their self-reported withdrawal symptoms. Regardless of patient characteristics, the treatment plan includes a dose increase of 4/1 mg/day every two-weeks until a maximum dose of 24/6 mg/day is reached. 
    \end{enumerate} & Dynamic \\
    \midrule
    \begin{enumerate}
        \setcounter{enumi}{3}
        \item All patients are prescribed BUP-NX at an initial visit. All patients receive an initial dose of 4/1 mg/day. At each two-week follow-up visit, patients who report any opioid use since the most recent visit receive a 4/1 mg/day BUP-NX dose increase, up to a maximum dose of 24/6 mg/day; patients who do not report any opioid use remain at the most recently prescried dose. 
        \end{enumerate} & Dynamic \\
    \bottomrule
    \end{tabular}
    \label{tab:rule_examples}
\end{table}
		
\subsection*{Optimal dynamic treatment rules} 
	
In any particular application, dependent on the number of covariates and their possible values, there are numerous potential dynamic treatment rules. We will denote the set of all possible treatment rules as $\mathcal{D}$. One of these treatment rules $\d$ in the set $\mathcal{D}$ is \emph{optimal} in the sense that it will maximize the benefit of a treatment in the population. \citep{murphyODTR2003} We can define a treatment as ``beneficial'' for a particular subgroup characterized by covariate values, $V=v$, if their---potentially unobserved---counterfactual outcome of interest had they received treatment is greater (if the outcome is desirable; or smaller, if the outcome is undesirable) than if they hadn't received treatment.

If we consider the expected value of the counterfactual outcome under a given treatment rule, $\E[Y_\d]$, it then follows that the ODTR is the dynamic rule, $\d_{\text{opt}}$, that of all possible rules, $\mathcal{D}$, maximizes (for desirable outcomes; or minimizes, for harmful outcomes) $\E[Y_\d]$, which can formally be denoted:
	
$$
\d_{\text{opt}} \equiv \argmax_{\d \in \mathcal{D}} \E [Y_{\d}].
$$
In the context of OUD research, estimating the ODTR would be relevant for research questions such as, ``using information collected at intake, for which patients should we increase BUP-NX dosage after one-month of treatment to minimize week-12 return-to-regular-opioid-use?''.

When a treatment rule may vary over time, the time-varying ODTR is the sequence of treatment rules, $\mathbf{D}_{\text{opt}}$, that maximizes (or minimizes) the expected value of the counterfactual outcome under that sequence of rules over all time-points, $\E [Y_{\mathbf{D}}]$, which can formally be denoted:

$$
\mathbf{D}_{\text{opt}} \equiv \argmax_{\mathbf{D} \in \mathcal{D}} \E [Y_{\mathbf{D}}].
$$
In the context of OUD research, estimating the time-varying ODTR would be relevant for research questions such as, ``using each patient's baseline history and information obtained in their weekly follow-up visits, for which patients should we increase dosage of BUP-NX each week to minimize week-12 return-to-regular-opioid-use?''

\subsection*{Evaluating the expected outcome under the ODTR}\label{section:psi} 
 
After learning an ODTR, it is often of interest to estimate the expected counterfactual outcome had the ODTR been applied. This target causal parameter is denoted $\psi \equiv \E[Y_{\d_{\text{opt}}}]$ when treatment is fixed in time or $\psi \equiv\E[Y_{\mathbf{D_\text{opt}}}]$ when treatment may vary over time. It is interpreted as the expected value of the counterfactual outcome under the optimal dynamic treatment rule---the average value of the counterfactual outcome of interest if, possibly contrary to what was observed, treatment had been assigned according to the optimal rule.

\section*{Estimation}\label{estimation}

Multiple strategies exist for estimating the ODTR; in general, approaches fall into two broad categories: methods based on regression of the conditional average treatment effect (CATE; most commonly Q-learning \citep{watkins1989learning,moodieQlearning2012} and A-learning \citep{murphyODTR2003}), and classification methods that directly estimate the ODTR (outcome weighted learning [OWL] \citep{owl2012}, residual weighted learning [RWL] \citep{ZhouRWL}, and augmented outcome weighted learning \citep{YingAOWL2018}). In this paper, we described a nonparametric doubly-robust regression-based strategy to learn the ODTR. 

\subsection*{Notation}
We first give some further notation that is necessary for describing how to estimate the longitudinal ODTR. Formally, consider a discrete-time process, $t = 1, ..., \tau$, where we observe $n$ i.i.d. copies $O_1, ..., O_n$ of $O = (L_1, A_1,\ldots,L_\tau,A_\tau, Y)$ generated from some unknown joint probability distribution $P$. At each time-point $t$, covariates $L_t$ precede a binary treatment indicator $A_t$ and an outcome $Y$. We use $H_t = (L_1, ..., L_t, A_1, ..., A_{t-1})$ to denote the history of all variables up until just before $A_t$. As previously mentioned, $V_t$ is the subset of covariates $H_t$ that will be used in the rule for assigning $A_t$. Capital letters (e.g., $A$) indicate random variables while lower-case letters (e.g., $a$) indicate realizations of a random variable. Symbols with a subscript $n$ represent values estimated from data (e.g., $P_n$ is an estimate of $P$). 

\subsection*{The blip function and DR-estimation of the ODTR for a single time-point}

The conditional average treatment effect (CATE), commonly referred to as the blip function\citep{robins2004optimal}, is a key component of learning the optimal rule using regression-based approaches.\citep{van2015targeted} Using, first, causal notation, and second, notation as a function of the observed data, the blip function is defined as: 

\begin{equation}\label{blip}
  B(V) = \E[Y_1 - Y_0 | V] \equiv \E\Big[\E[Y \mid H, A = 1] - \E[Y \mid H, A = 0] \Big| V\Big].
\end{equation}
Assuming the counterfactual outcome for an observation under their observed treatment level is the value of the outcome that we did actually observe (consistency), that all common causes of the exposure and outcome are measured and adjusted for (no-unmeasured confounding or exchangeability), and that the probability of experiencing all treatment levels is non-zero for all observations (positivity), the two are equal \citep{van2006statistical}.
The optimal treatment for a given observation is assigned according to the sign of the estimated blip, $B_n(V)$. For example, if we wish to maximize $\E [Y_\d]$, then we would assign treatment to individuals where $B_n(V)$ is positive and not assign treatment where $B_n(V)$ is negative: 

\begin{equation}\label{rule}
  \d_{\text{opt}}(V) \equiv \mathrm{I}(B(V) > 0).
\end{equation}
Equation \ref{blip} reveals a simple substitution estimator for $\d_{\text{opt}}(V)$. We can first estimate $Q(H, A) = \E[Y \mid H, A]$. We can then obtain estimates of the CATE: $Q(H, A = 1) - Q(H, A = 0)$ by predicting $Q_n(H, A)$ where $A$ is fixed to $1$ and $0$ respectively. The estimate $B_n(V)$ is then given by regressing $Q_n(H, A = 1) - Q_n(H, A = 0)$ on $V$ and then predicting from this model with $V$ equal to the values observed ($V=v$). 

While simple, performance of this estimator can be poor,\cite{kennedy2020towards} and we can improve upon this strategy by means of an \emph{unbiased transformation} of $B(V)$. A function $D(O)$ of the data is said to be an unbiased transformation if $\E[D(O) \mid V] = B(V)$. We focus on the augmented inverse probability weighting (AIPW)\citep{robins1994estimation} transformation of the ATE:

\begin{equation}\label{eif1}
  D(Q,g)(O) = \frac{2A - 1}{g(A,H)} \Big( Y - Q(H, A) \Big) + Q(H, A = 1) - Q(H, A = 0),  
\end{equation}
where $g(A,H) = \Pr(A = a \mid H)$. This is an improvement, because the AIPW transformation is \emph{doubly-robust}; it requires, in addition to the estimation of $Q(H, A)$, an estimate for the probability of receiving the observed treatment $g(A,H)$. If either $Q(H, A)$ or $g(A,H)$ is consistently estimated we may obtain an unbiased estimate of $B(V)$. The double robustness property also allows us to use flexible, data-adaptive algorithms for model fitting and make inferences about our estimates using standard statistical theory.

Pseudo code for this algorithm is presented in Algorithm \ref{alg:odtr}.

\begin{algorithm}[H]
\caption{Estimation of $\d_{\text{opt}}$}\label{alg:odtr}
\begin{algorithmic}[1]
\State $\m \gets$ \texttt{Regress}$(\text{outcome} = Y, \text{predictors} = \{H, A\})$
\State $\g \gets$ \texttt{Regress}$(\text{outcome} = A, \text{predictors} = H)$
\State $Q_n(H, A = 1) \gets$ \texttt{Predict}$(\text{model} = \m, \text{predictors} = \{H, A = 1\})$
\State $Q_n(H, A = 0) \gets$ \texttt{Predict}$(\text{model} = \m, \text{predictors} = \{H, A = 0\})$
\State $g_n \gets$ $A \times$ \texttt{Predict}$(\g, H) + (1 - A) \times (1 - $ \texttt{Predict}$(\g, H))$
\State $D_n(Q, g) \gets$ \texttt{AIPW\_Transformation}$(Y, Q_n(H, A = 1), Q_n(H, A = 0), g_n)$ \Comment{Equation \ref{eif1}}
\State $\Tilde{\m} \gets$ \texttt{Regress}$(\text{outcome} = D_n(Q, g), \text{predictors} = V)$
\State $\Tilde{B}_n(V) \gets$ \texttt{Predict}$(\text{model} = \Tilde{\m}, \text{predictors} = V)$
\State $\d_{n, \text{opt}}(V) \gets \mathrm{I}(\Tilde{B}_n(V) > 0)$ 
\State \Return $\d_{n, \text{opt}}(V)$ 
\end{algorithmic}
\end{algorithm}

\subsection*{Extension of the DR-estimator to allow the ODTR to change over time}

How can we move from estimating a single rule to estimating a sequence of rules? Consider a study with two time-points. The ODTR, $\mathbf{D}_{\text{opt}}$, is the sequence of rules $[\d_1, \d_2]$ that maximizes $\E[Y_{A_1 = \d_1, A_2 = \d_2}]$. 

All rules must have support in the data. This means that we must observe observations that follow each rule. Intuitively, this leads us to begin at the last time-point, where we maximize the expected value of the counterfactual outcome under a treatment rule at the last time-point. The optimal rule $\d_{2, \text{opt}}$ is the rule that maximizes $\E[Y_{A_2 = \d_2}]$. This is the same estimation problem as already discussed and thus we can use Algorithm \ref{alg:odtr} to generate an estimate of $\d_{2, \text{opt}}$. With an estimate of $ \d_{2, \text{opt}}$, our goal of estimating $\mathbf{D}_{\text{opt}}$ has now changed from finding $[\d_1, \d_2]$ to maximize $\E[Y_{A_1 = \d_1, A_2 = \d_2}]$ to finding $\d_1$ to maximize $\E[Y_{A_1 = \d_1, A_2 = \d_{2, \text{opt}}}]$.

The optimal treatment rule for treatment at the first time-point is then the rule that maximizes the expected value of the counterfactual outcome in a world where treatment at the second time-point has already been optimally assigned. To solve this, we must redefine the blip function:

\begin{equation*}
    \begin{split}
        B_1(V) &= \E[Y_{A_1 = 1, A_2 = \d_{2, \text{opt}}} - Y_{A_1 = 0, A_2 = \d_{2, \text{opt}}} \mid V_1] \\
     &= \E\bigg[\E\Big(\E[Y \mid H_2, A_2 = \d_{2, \text{opt}}] \Bigm| H_1, A_1 = 1 \Big) - \E\Big(\E[Y \mid H_2, A_2 = \d_{2, \text{opt}}] \Bigm| H_1, A_1 = 0 \Big) \biggm| V_1 \bigg].
\end{split}
\end{equation*}
We can estimate $\E(\E[Y \mid H_2, A_2 = \d_{2, \text{opt}}] \mid H_1, A_1)$ by regressing $\Tilde{Y} = Q_2(H_2, A_2 = \d_{2, \text{opt}})$, the expected value of $Q_2$ when $A_2$ is set to $\d_{2, \text{opt}}$, 
on $H_1$ and $A_1$. The optimal rule for $A_1$ is then given by $\d_{1, \text{opt}} = \mathrm{I}(B_1(V) > 0)$, and $\mathbf{D}_{\text{opt}} = [\d_{1, \text{opt}}, \d_{2, \text{opt}}]$.

We can generalize this procedure to any time-horizon, $\tau$, and also apply the strategy of using doubly-robust unbiased transformations of $B_t(V)$. Pseudo code for this algorithm is presented in Algorithm \ref{alg:lodtr}. First we amend the notation for the result of the AIPW transformation\citep{luedtkeSLodtr} (\ref{eif3}) as $D_t(\d_{t+1, \text{opt}}, Q, g)$ to highlight that the optimal rule at the current time is defined with respect to the distribution of the counterfactual outcome if at the next time-point treatment was optimally assigned:

\begin{equation}\label{eif3}
\begin{split}
    D_t(\d_{t+1, \text{opt}}, Q, g)(O) = &\sum_{s = t}^{\tau} \bigg\{ \frac{2A_t - 1}{g_t(A_t, H_t)} \prod_{k = t+1}^{s} \frac{\mathds{1}(A_{k} = \d_{k, \text{opt}})}{g_{k}(A_{k}, H_{k})} \bigg\} \times \Big(Q_{s+1}(H_{s+1}, \d_{s+1,\text{}}) - Q_s(H_s, A_s)\Big)\\ &+ Q_t(H_t, A_t = 1) - Q_t(H_t, A_t = 0).
\end{split}
\end{equation}
Note that when $t = \tau$ or when $\tau = 1$, $D_t(\d_{t+1, \text{opt}}, Q, g)$ reduces to Equation \ref{eif1}.

\begin{algorithm}[H]
\caption{Generalization of Algorithm \ref{alg:odtr} to estimate $\mathbf{D}_{ \text{opt}}$}\label{alg:lodtr}
\begin{algorithmic}[1]
\State $\Tilde{Y} \gets Y$
\For {$t = \tau, \dots, 1$}
\State $\m \gets$ \texttt{Regress}$(\text{outcome} = \Tilde{Y}, \text{predictors} = \{H_t, A_t\})$
\State $\g \gets$ \texttt{Regress}$(\text{outcome} = A_t, \text{predictors} = H_t)$
\State $Q_n(H_t, A_t = 1) \gets$ \texttt{Predict}$(\text{model} = \m, \text{predictors} = \{H_t, A_t = 1\})$
\State $Q_n(H_t, A_t = 0) \gets$ \texttt{Predict}$(\text{model} = \m, \text{predictors} = \{H_t, A_t = 0\})$
\State $g_{n, t} \gets$ $A_t \times$ \texttt{Predict}$(\text{model} = \g, \text{predictors} = H_t)\newline
        \hspace*{4em} + (1 - A_t) \times (1 - $ \texttt{Predict}$(\text{model} = \g, \text{predictors} = H_t))$
\State $D_{n, t}(Q, g) \gets$ \texttt{AIPW\_Transformation}$(Y, Q_n(H_t, A_t = 1), Q_n(H_t, A_t = 0), g_{n, t})$ \Comment{Equation \ref{eif3}}
\State $\Tilde{\m} \gets$ \texttt{Regress}$(\text{outcome} = D_{n, t}(Q, g), \text{predictors} = V_t)$
\State $\Tilde{B}_{n, t}(V_t) \gets$ \texttt{Predict}$(\text{model} = \Tilde{\m}, \text{predictors} = V_t)$
\State $\d_{n, t, \text{opt}} \gets \mathrm{I}(\Tilde{B}_{n, t}(V_t) > 0)$ 
\State $\Tilde{Y} \gets$ \texttt{Predict}$(\text{model} = \m, \text{predictors} = \{H_t, A = \d_{n, t, \text{opt}})\}$
\EndFor
\State $\mathbf{D}_{n, \text{opt}} \gets [\d_{n, 1, \text{opt}}, \dots, \d_{n, \tau, \text{opt}}]$
\State \Return $\mathbf{D}_{n, \text{opt}}$
\end{algorithmic}
\end{algorithm}

The regressions in the above estimation algorithms can be fitted by data-adaptive, machine-learning algorithms, allowing for flexible modelling while maintaining valid statistical inference. One could use, for example, a decision tree to estimate $\Tilde{B}(V)$ and generate interpretable decision rules. 

An \texttt{R} package to implement the method we describe is available for download from \href{https://github.com/Rudolph-et-al-MSPH-Epidemiology/odtr}{GitHub}. We conducted a limited simulation study to evaluate correct implementation of the estimator. However, we acknowledge that these simulations are not informative of general estimator performance. Results of the simulation are presented in the Supplementary Materials. 

\subsection*{Approaches to estimating the expected counterfactual outcome value under the ODTR}

The mean counterfactual outcome had the ODTR been applied can be estimated using different techniques such as targeted minimum loss-based estimation \citep{tmleLong, van2015targeted} or estimators based on doubly-robust unbiased transformations \citep{luedtke2018sequential}. In addition to the standard set of assumptions for identifying causal estimands from observational data, we refer readers to \citet{van2015targeted} for technical details on the necessary conditions required for consistently (i.e., approximately ``unbiased'' as sample size increases) estimating the mean outcome under the learned ODTR and its confidence intervals. Intuitively, these assumptions include the assumption that the outcome regressions and propensity scores are estimated correctly, as well as the assumption that the estimated functions are ``smooth enough'', although this assumption can be bypassed using sample splitting.\cite{van2011cross} Importantly, the assumptions also include an assumption that the blip function is non-zero, that is, $P(B(V)=0)=0$, so that no unit has a conditional average treatment effect of exactly zero. 

\section*{Example analysis: Learning the ODTR of when to increase BUP-NX dose to minimize RROU}

Learning an ODTR for BUP-NX dosage is of interest, because, currently, dosing decisions are made at the discretion of addiction medicine clinicians and vary substantially, even among clinicians given uniform guidance as part of clinical trials \citep{rudolph2022dose}. Under-dosing of medication used in treating OUD, like BUP-NX, is relatively common and considered a contributing factor to poor OUD treatment outcomes. \citep{gordon2015patterns,d2019evidence} Recently, \citet{rudolph2022dose} examined a time-varying, clinically-defined dosing strategy where BUP-NX dose was increased whenever a patient indicated opioid use in the prior week, and found that this strategy resulted in a 7\% reduced risk of RROU by week 12 of treatment compared to a treatment strategy where BUP-NX dose was held constant. It is of interest to know whether an ODTR learned from the data could reduce risk of RROU even further. 

We used the DR-estimator described in Section 3.3 to learn a time-varying ODTR for when to increase BUP-NX dose during the first 6-weeks of treatment to minimize RROU. We compared the estimated expected value of RROU under the learned ODTR to the estimated expected value of RROU under two clinically defined dosing strategies: 1) holding dose constant after three-weeks of treatment (\emph{constant dosing rule}; $\d_0$); and 2) increasing dose in response to opioid use during the previous week (\emph{dynamic dosing rule}; $\d_1$). Note that $\d_1$ was prespecified, is only a function of opioid use at the prior week, and may not be optimal. In comparison, $\mathbf{D}_{opt}$ is learned, is a function of all baseline and time-varying covariates, and is optimal.

We used data from the BUP-NX arms of three harmonized comparative effectiveness trials of medications for the treatment of OUD that were part of the NIDA Clinical Trials Network, $n = 2,199$ patients.\citep{potter2013buprenorphine,saxon2013buprenorphine,weiss2010multi,lee2018comparative} Details of the treatment arms, sample size, and study years for each trial in the harmonized cohort is shown in Table \ref{tab:t0}. 

\begin{table}[H]
\centering\footnotesize
\caption{Individual Clinical Trial Network (CTN) studies comprising the harmonized BUP-NX study cohort.}
\begin{tabular}{lrrr}
\toprule
Study &  CTN0027 & CTN0030 Phase II & CTN0051 \\
			\midrule
			Study years & 2006-2010 & 2006-2009 & 2014-2017 \\
			Sample size & 1269 & 360 & 570 \\
			Study length (weeks) & 24 & 12 & 24 \\
			Alternate arm$^a$  & methadone & BUP-NX/drug counseling & XR-NTX$^b$ \\
			\bottomrule
\multicolumn{4}{l}{\footnotesize{$a$ Subjects were randomized to receive this treatment arm or BUP-NX with standard medical care.}}\\
\multicolumn{4}{l}{\footnotesize{$b$ Extended-release naltrexone.}}\\
\end{tabular}
\label{tab:t0}
\end{table}

We used RROU as the outcome of interest, defined as the last day of 7-days of non-study opioid use, or the first-day of the fourth consecutive week of at-least-once-weekly non-study opioid use. Non-study opioid use was determined using urine drug screens and timeline follow-back interviews. Missed visits and urine screening refusal were considered a positive for opioid use, consistent with the primary trial papers and previous secondary analyses, \citep{lee2018comparative,weiss2011adjunctive,potter2013buprenorphine,rudolph2022dose,rudolph2022optimally} and is considered a reasonable assumption given prior research. In all three trials, BUP-NX was dispensed for self-administration with dose and dose adjustments prescribed according to clinical judgment. Our time-varying treatment of interest was an increase in the maximum weekly dispensed BUP-NX dose from the prior-week, a binary (0/1) variable. \citep{rudolph2022dose} 

Baseline confounders of the time-varying BUP-NX dose increase and time to RROU relationship included race/ethnicity, age, biological sex, highest level of past opioid withdrawal discomfort, past-year substance use disorders, history of neurological injury, history of psychiatric disorders, history of IV drug use, and past-30 day drug use. Time-varying confounders included most recently prescribed BUP-NX dose and an indicator of non-study opioid use during the prior week.

Cross-fit (CF) estimates of $Q_t(H_t, A_t)$ and $g_t(H_t, a_t)$ were fit using the discrete super learner \cite{van2007super} with a learner library that included $\ell$-1 regularization \cite{tibshirani1996regression}, multivariate adaptive regression splines [MARS] \cite{friedman1991multivariate}, and gradient boosting.\cite{lightgbm} Super learning combines predictions from a set of user-defined \textit{learners} using another algorithm, called the meta-learner. \citep{breimanStacked,van2007super} We used a "winner-take-all" meta-learner, with the winner (the discrete super learner) chosen as the learner with the best cross-validated mean squared error. For a full explanation of the super learner, we refer the reader to \citet{van2007super} and \citet{phillips2023practical}. Cross-fit estimates of $\Tilde{B}_t(V)$ were fit using $\ell$-1 regularization. 

We used a longitudinal sequentially doubly robust \citep{luedtke2018sequential} (SDR) estimator to estimate the mean counterfactual week-6 relapse under each dosing strategy, $\psi \equiv [\E(Y_{\d0})$, $\E(Y_{\d1})$, and $\E[Y_{\textbf{D}_{n, \text{opt}}})$]. The SDR estimator incorporated the same super learner library that was used for fitting $Q_t(H_t, A_t)$ and $g_t(H_t, a_t)$. We summarize our results using risk ratios (RR), comparing the mean counterfactual week-6 RROU from the learned ODTR and the dosing rule from \citet{rudolph2022dose} to the constant dose rule. Variances were estimated using the sample variance of the influence function of the $\log$ RR.  \cite{luedtke2018sequential} We used \texttt{R}\cite{R} (version 4.2.1) for all analyses with the lmtp \cite{lmtpR, lmtpJASA} packages. Code to replicate the analyses is available on \href{https://github.com/Rudolph-et-al-MSPH-Epidemiology/odtr}{GitHub}.

\subsection*{Results}

The number of patients who would have BUP-NX dose increased or held constant and were observed as having naturally followed a given rule are shown in Table \ref{tab:results}. Table \ref{tab:results} also provides the estimated expected risk of week-6 RROU under the dosing rules. Under the rule of increasing dose in response to prior-week opioid use, $\d1$, many individuals would receive dose increases early in treatment, but fewer as treatment progresses. This is because the likelihood of using opioids decreases in time as BUP-NX treatment continues. In contrast, under the learned ODTR (without cross-fitting), BUP-NX dose would be increased for all patients every week of treatment, excluding week three. Inspecting the models for $\Tilde{B}_t(V)$ revealed that none of the variables in $V_t$ were retained as selected features and that the models reduced to intercept-only models with the optimal rule for a given week corresponding to the sign of $D_{n, t}(Q, g)$. 

We find evidence that the learned ODTR would reduce the risk of RROU compared to holding BUP-NX dose constant by 16.6\% ($\hat{\text{RR}}$: 0.83, 95\% CI: 0.68, 1.02). Applying the dynamic dosing rule, $\d1$, from \citet{rudolph2022dose}, we estimated a 10.5\% ($\hat{\text{RR}}$: 0.90, 95\% CI: 0.84, 0.96) reduction in risk of RROU compared to holding dose constant. Comparing the learned ODTR to the dynamic dosing rule, we estimated a 6.8\% reduction in risk off RROU, however this difference was not meaningful ($\hat{\text{RR}}$: 0.93, 95\% CI = 0.77, 1.13).

\begin{table}[H]
\caption{Number of patients randomized to receive BUP-NX that were observed as following a given rule: (Increase) increased dose under the rule and were observed as increasing dose, or (Constant) had a constant dose under the rule and were observed as having a constant dose; and, estimates of the expected counterfactual week-6 RROU under the given rule ($\hat\psi$).}
\centering\footnotesize
\begin{tabular}[t]{llrrrrc}
\toprule
& & Wk. 2 & 3 & 4 & 5 & $\hat\psi \equiv \E[Y_{\mathbf{D}}]$\\
\midrule
\multicolumn{7}{l}{Learned ODTR, $\mathbf{D}_{n, \text{opt}}$}\\
\midrule
\hspace{1em}\multirow{3}{8em}{Based on the CF ODTR} & Increase & 236 & 30 & 84 & 44 & \multirow{2}{*}{0.324}\\
\hspace{1em} & Constant & 319 & 888 & 292 & 285 & \multirow{2}{*}{95\% CI = 0.261,0.387}\\
\hspace{1em}& \emph{Total} & 555 & 918 & 376 & 329\\[2pt]
\hdashline\noalign{\vskip 1ex}
\hspace{1em}\multirow{3}{8em}{Based on the ODTR learned without CF} & Increase & 320 & 0 & 118 & 69 & \\
\hspace{1em}& Constant & 0 & 1109 & 0 & 0 & \\
\hspace{1em}& \emph{Total} & 320 & 1109 & 118 & 69 \\
\addlinespace[0.3em]
\midrule
\addlinespace[0.3em]
\multicolumn{7}{l}{Constant dosing rule}\\
\midrule
\hspace{1em}& \emph{Total} & 1028 & 1109 & 947 & 928 & \multirow{2}{*}{0.388}\\
\hspace{1em}& \hspace{1em}$\mathbf{D}_{n, \text{opt}}\,^a$ & 675 & 180 & 611 & 545 & \multirow{2}{*}{95\% CI = 0.358, 0.419}\\
\hspace{1em}& \hspace{1em}$\d1\,^b$ & 375 & 269 & 227 & 125\\
\addlinespace[0.3em]
\midrule
\multicolumn{7}{l}{\citet{rudolph2022dose} dynamic dosing rule, $\d_1$}\\
\midrule
\hspace{1em}& Increase & 147 & 59 & 40 & 24 & \multirow{2}{*}{0.348}\\
\hspace{1em}& Constant & 605 & 702 & 685 & 698 & \multirow{2}{*}{95\% CI = 0.327, 0.369}\\
\hspace{1em}& \emph{Total} & 752 & 761 & 725 & 722\\
\bottomrule \\
\multicolumn{7}{l}{\footnotesize{$^a$ No. of dose increases that would have been indicated applying $\mathbf{D}_{n, \text{opt}}$}}\\
\multicolumn{7}{l}{\footnotesize{$^b$ No. of dose increases that would have been indicated applying $\d1$}}\\
\label{tab:results}
\end{tabular}
\end{table}

\section*{Discussion}

The average treatment effect is a limited causal effect measure insofar as it ignores treatment effect heterogeneity that may be practically important for improving treatment success for certain subgroups of patients. While state-of-the-art methods exist for quantifying treatment effect heterogeneity and for estimating optimal treatment rules,\citep{watkins1989learning,moodieQlearning2012,murphyODTR2003,owl2012,ZhouRWL,YingAOWL2018,luedtkeSLodtr,MontoyavanderLaanLuedtkeSkeemCoylePetersen2022, DiazEnsembleTTE,zhaoBowl} the papers establishing the theory for these methods are not always approachable for the applied analyst. 

Here, we provided intuition for dynamic treatment regimes and optimal dynamic treatment regimes and presented a doubly-robust and flexible modeling algorithm for estimating the ODTR for both single time-point and time-varying treatments. We applied this algorithm to learn a longitudinal ODTR to minimize RROU over 6-weeks of BUP-NX MOUD treatment. We found evidence that the learned optimal rule would decrease the risk of week-6 RROU compared to a strategy where BUP-NX dose is held constant for the duration of treatment, and compared similarly well to a pre-specified dynamic dosing strategy where BUP-NX dose was increased in response to opioid use in the prior week.

The learned ODTR resulted in a treatment strategy where BUP-NX dose would be increased at weeks two, four, and five of treatment, irrespective of patient history. This ODTR is not particularly surprising given that: 1) under-dosing of BUP-NX is common,\citep{gordon2015patterns} 2) a key contributor to poor OUD treatment outcomes,\citep{d2019evidence} and 3) we considered only the first six weeks of treatment, during which time evidence suggests that dose should be increased regularly until a minimum threshold (e.g., 16 mg \citep{tip2021}) is reached.\citep{rudolph2022dose} We used a relatively simple model to estimate the ODTR, however, any regression technique in the machine learning literature could be used, including more flexible, data-adaptive algorithms.

While we focused on presenting a single approach for learning the longitudinal ODTR, other methods exist. Assuming their respective assumptions are satisfied, all of the algorithms previously mentioned in the Estimation Section are asymptotically consistent estimators of the ODTR; however, their finite sample performance may vary. As a result, one algorithm may perform better than another in any particular analysis.\citep{WangEHR} It is also plausible that the optimal treatment strategy is a simple, static treatment rule, as may be the case with our illustrative application. With this in mind, another approach, which we refer to as ODTR \emph{ensembles} ,\citep{luedtkeSLodtr,MontoyavanderLaanLuedtkeSkeemCoylePetersen2022, DiazEnsembleTTE} allows the analyst to specify a set of ODTR estimation algorithms (e.g., including both regression- and classification-based methods together with simple, static interventions) to ``compete'' against each other in estimating the ODTR, with the optimal strategy data-adaptively chosen. We also limited our discussion to learning the optimal decision rule for binary treatments and a single outcome. However, the described approach can be modified to handle categorical exposures,\citep{tlverseHandbook} resource constrained settings,\citep{luedtke2016optimal} informative-censoring,\citep{luedtkeSLodtr} and survival outcomes.\citep{DiazEnsembleTTE}

\newpage

\newpage
\renewcommand{\refname}{References}

\bibliography{lib}

\begin{thebibliography}{46}
\providecommand{\natexlab}[1]{#1}
\providecommand{\url}[1]{\texttt{#1}}
\expandafter\ifx\csname urlstyle\endcsname\relax
  \providecommand{\doi}[1]{doi: #1}\else
  \providecommand{\doi}{doi: \begingroup \urlstyle{rm}\Url}\fi

\bibitem[Abuse et~al.(2021)]{tip2021}
Substance Abuse et~al.
\newblock Medications for opioid use disorder for healthcare and addiction
  professionals, policymakers, patients, and families: Treatment improvement
  protocol tip 63.
\newblock 2021.
\newblock URL
  \url{https://store.samhsa.gov/sites/default/files/SAMHSA_Digital_Download/PEP21-02-01-002.pdf}.

\bibitem[Breiman(1996)]{breimanStacked}
Leo Breiman.
\newblock Stacked regressions.
\newblock \emph{Machine Learning}, 24, 1996.
\newblock \doi{10.1007/BF00117832}.

\bibitem[Cain et~al.(2010)Cain, Robins, Lanoy, Logan, Costagliola, and
  Hern{\'a}n]{cain2010start}
Lauren~E Cain, James~M Robins, Emilie Lanoy, Roger Logan, Dominique
  Costagliola, and Miguel~A Hern{\'a}n.
\newblock When to start treatment? a systematic approach to the comparison of
  dynamic regimes using observational data.
\newblock \emph{The international journal of biostatistics}, 6\penalty0 (2),
  2010.

\bibitem[D'Aunno et~al.(2019)D'Aunno, Park, and Pollack]{d2019evidence}
Thomas D'Aunno, Sunggeun~Ethan Park, and Harold~A Pollack.
\newblock Evidence-based treatment for opioid use disorders: A national study
  of methadone dose levels, 2011--2017.
\newblock \emph{Journal of Substance Abuse Treatment}, 96, 2019.

\bibitem[Díaz et~al.(2018)Díaz, Savenkov, and Ballman]{DiazEnsembleTTE}
Iván Díaz, Oleksandr Savenkov, and Karla Ballman.
\newblock {Targeted learning ensembles for optimal individualized treatment
  rules with time-to-event outcomes}.
\newblock \emph{Biometrika}, 105\penalty0 (3):\penalty0 723--738, 2018.
\newblock \doi{10.1093/biomet/asy017}.

\bibitem[Díaz et~al.(2021)Díaz, Williams, Hoffman, and Schneck]{lmtpJASA}
Iván Díaz, Nicholas Williams, Katherine Hoffman, and Edward Schneck.
\newblock Non-parametric causal effects based on longitudinal modified
  treatment policies.
\newblock \emph{Journal of the American Statistical Association}, 2021.
\newblock \doi{10.1080/01621459.2021.1955691}.

\bibitem[Friedman(1991)]{friedman1991multivariate}
Jerome~H Friedman.
\newblock Multivariate adaptive regression splines.
\newblock \emph{The Annals of Statistics}, pages 1--67, 1991.

\bibitem[Gordon et~al.(2015)Gordon, Lo-Ciganic, Cochran, Gellad, Cathers,
  Kelley, and Donohue]{gordon2015patterns}
Adam~J Gordon, Wei-Hsuan Lo-Ciganic, Gerald Cochran, Walid~F Gellad, Terri
  Cathers, David Kelley, and Julie~M Donohue.
\newblock Patterns and quality of buprenorphine opioid agonist treatment in a
  large medicaid program.
\newblock \emph{Journal of Addiction Medicine}, 9\penalty0 (6), 2015.

\bibitem[Hernán and Robins(2020)]{hernanWhatIfDynamic}
Miguel~A Hernán and James~M Robins.
\newblock \emph{Causal Inference: What If}, chapter Time-varying treatments.
\newblock Chapman \& Hall/CRC, Boca Raton, 2020.

\bibitem[Kennedy(2020)]{kennedy2020towards}
Edward~H Kennedy.
\newblock Towards optimal doubly robust estimation of heterogeneous causal
  effects.
\newblock \emph{arXiv preprint arXiv:2004.14497}, 2020.

\bibitem[Kosorok and Laber(2019)]{kosorok2019precision}
Michael~R Kosorok and Eric~B Laber.
\newblock Precision medicine.
\newblock \emph{Annual review of statistics and its application}, 6:\penalty0
  263--286, 2019.

\bibitem[Kosorok et~al.(2021)Kosorok, Laber, Small, and
  Zeng]{kosorok2021introduction}
Michael~R Kosorok, Eric~B Laber, Dylan~S Small, and Donglin Zeng.
\newblock Introduction to the theory and methods special issue on precision
  medicine and individualized policy discovery, 2021.

\bibitem[Lee et~al.(2018)Lee, Nunes~Jr, Novo, Bachrach, Bailey, Bhatt, Farkas,
  Fishman, Gauthier, Hodgkins, et~al.]{lee2018comparative}
Joshua~D Lee, Edward~V Nunes~Jr, Patricia Novo, Ken Bachrach, Genie~L Bailey,
  Snehal Bhatt, Sarah Farkas, Marc Fishman, Phoebe Gauthier, Candace~C
  Hodgkins, et~al.
\newblock Comparative effectiveness of extended-release naltrexone versus
  buprenorphine-naloxone for opioid relapse prevention ({X:BOT}): a
  multicentre, open-label, randomised controlled trial.
\newblock \emph{The Lancet}, 391\penalty0 (10118):\penalty0 309--318, 2018.

\bibitem[Liu et~al.(2018)Liu, Wang, Kosorok, Zhao, and Zeng]{YingAOWL2018}
Ying Liu, Yuanjia Wang, Michael~R. Kosorok, Yingqi Zhao, and Donglin Zeng.
\newblock Augmented outcome-weighted learning for estimating optimal dynamic
  treatment regimens.
\newblock \emph{Statistics in Medicine}, 37\penalty0 (26):\penalty0 3776--3788,
  2018.
\newblock \doi{https://doi.org/10.1002/sim.7844}.

\bibitem[Luedtke and {van der Laan}(2016{\natexlab{a}})]{luedtkeSLodtr}
Alexander Luedtke and Mark {van der Laan}.
\newblock Super-learning of an optimal dynamic treatment rule.
\newblock \emph{The International Journal of Biostatistics}, 12:\penalty0
  305--332, 05 2016{\natexlab{a}}.
\newblock \doi{10.1515/ijb-2015-0052}.

\bibitem[Luedtke and {van der Laan}(2016{\natexlab{b}})]{luedtke2016optimal}
Alexander~R Luedtke and Mark~J {van der Laan}.
\newblock Optimal individualized treatments in resource-limited settings.
\newblock \emph{The International Journal of Biostatistics}, 12\penalty0
  (1):\penalty0 283--303, 2016{\natexlab{b}}.

\bibitem[Luedtke et~al.(2018)Luedtke, Sofrygin, {van der Laan}, and
  Carone]{luedtke2018sequential}
Alexander~R. Luedtke, Oleg Sofrygin, Mark~J. {van der Laan}, and Marco Carone.
\newblock Sequential double robustness in right-censored longitudinal models,
  2018.

\bibitem[Montoya et~al.(2022)Montoya, {van der Laan}, Luedtke, Skeem, Coyle,
  and Petersen]{MontoyavanderLaanLuedtkeSkeemCoylePetersen2022}
Lina~M. Montoya, Mark~J. {van der Laan}, Alexander~R. Luedtke, Jennifer~L.
  Skeem, Jeremy~R. Coyle, and Maya~L. Petersen.
\newblock The optimal dynamic treatment rule superlearner: considerations,
  performance, and application to criminal justice interventions.
\newblock \emph{The International Journal of Biostatistics}, 2022.
\newblock \doi{10.1515/ijb-2020-0127}.

\bibitem[Moodie et~al.(2012)Moodie, Chakraborty, and
  Kramer]{moodieQlearning2012}
Erica E.~M. Moodie, Bibhas Chakraborty, and Michael~S. Kramer.
\newblock Q-learning for estimating optimal dynamic treatment rules from
  observational data.
\newblock \emph{Canadian Journal of Statistics}, 40\penalty0 (4):\penalty0
  629--645, 2012.
\newblock \doi{https://doi.org/10.1002/cjs.11162}.

\bibitem[Murphy(2003)]{murphyODTR2003}
Susan~A Murphy.
\newblock Optimal dynamic treatment regimes.
\newblock \emph{Journal of the Royal Statistical Society: Series B (Statistical
  Methodology)}, 65\penalty0 (2):\penalty0 331--355, 2003.
\newblock \doi{https://doi.org/10.1111/1467-9868.00389}.

\bibitem[Phillips et~al.(2023)Phillips, van~der Laan, Lee, and
  Gruber]{phillips2023practical}
Rachael~V Phillips, Mark~J van~der Laan, Hana Lee, and Susan Gruber.
\newblock Practical considerations for specifying a super learner.
\newblock \emph{International Journal of Epidemiology}, 52\penalty0
  (4):\penalty0 1276--1285, 2023.

\bibitem[Potter et~al.(2013)Potter, Marino, Hillhouse, Nielsen, Wiest, Canamar,
  Martin, Ang, Baker, Saxon, et~al.]{potter2013buprenorphine}
Jennifer~S Potter, Elise~N Marino, Maureen~P Hillhouse, Suzanne Nielsen,
  Katharina Wiest, Catherine~P Canamar, Judith~A Martin, Alfonso Ang, Rachael
  Baker, Andrew~J Saxon, et~al.
\newblock Buprenorphine/naloxone and methadone maintenance treatment outcomes
  for opioid analgesic, heroin, and combined users: findings from starting
  treatment with agonist replacement therapies ({START}).
\newblock \emph{Journal of Studies on Alcohol and Drugs}, 74\penalty0
  (4):\penalty0 605--613, 2013.

\bibitem[{R Core Team}(2022)]{R}
{R Core Team}.
\newblock \emph{R: A Language and Environment for Statistical Computing}.
\newblock R Foundation for Statistical Computing, Vienna, Austria, 2022.
\newblock URL \url{https://www.R-project.org/}.

\bibitem[Robins(2004)]{robins2004optimal}
James~M Robins.
\newblock Optimal structural nested models for optimal sequential decisions.
\newblock In \emph{Proceedings of the Second Seattle Symposium in
  Biostatistics: analysis of correlated data}, pages 189--326. Springer, 2004.

\bibitem[Robins et~al.(1994)Robins, Rotnitzky, and Zhao]{robins1994estimation}
James~M Robins, Andrea Rotnitzky, and Lue~Ping Zhao.
\newblock Estimation of regression coefficients when some regressors are not
  always observed.
\newblock \emph{Journal of the American Statistical Association}, 89\penalty0
  (427):\penalty0 846--866, 1994.

\bibitem[Rudolph et~al.(2021)Rudolph, D{\'\i}az, Luo, Rotrosen, and
  Nunes]{rudolph2021optimizing}
Kara~E Rudolph, Iv{\'a}n D{\'\i}az, Sean~X Luo, John Rotrosen, and Edward~V
  Nunes.
\newblock Optimizing opioid use disorder treatment with naltrexone or
  buprenorphine.
\newblock \emph{Drug and Alcohol Dependence}, 228:\penalty0 109031, 2021.

\bibitem[Rudolph et~al.(2022{\natexlab{a}})Rudolph, Williams, Díaz, Luo,
  Rotrosen, and Nunes]{rudolph2022optimally}
Kara~E Rudolph, Nicholas~T Williams, Ivan Díaz, Sean~X Luo, John Rotrosen, and
  Edward~V Nunes.
\newblock Optimally choosing medication type for patients with opioid use
  disorder.
\newblock \emph{American Journal of Epidemiology}, 2022{\natexlab{a}}.

\bibitem[Rudolph et~al.(2022{\natexlab{b}})Rudolph, Williams, Goodwin, Shulman,
  Fishman, D{\'\i}az, Luo, Rotrosen, and Nunes]{rudolph2022dose}
Kara~E Rudolph, Nicholas~T Williams, Alicia T~Singham Goodwin, Matisyahu
  Shulman, Marc Fishman, Iv{\'a}n D{\'\i}az, Sean Luo, John Rotrosen, and
  Edward~V Nunes.
\newblock Buprenorphine \& methadone dosing strategies to reduce risk of
  relapse in the treatment of opioid use disorder.
\newblock \emph{Drug and alcohol dependence}, 239:\penalty0 109609,
  2022{\natexlab{b}}.

\bibitem[Saxon et~al.(2013)Saxon, Ling, Hillhouse, Thomas, Hasson, Ang,
  Doraimani, Tasissa, Lokhnygina, Leimberger, et~al.]{saxon2013buprenorphine}
Andrew~J Saxon, Walter Ling, Maureen Hillhouse, Christie Thomas, Albert Hasson,
  Alfonso Ang, Geetha Doraimani, Gudaye Tasissa, Yuliya Lokhnygina, Jeff
  Leimberger, et~al.
\newblock Buprenorphine/naloxone and methadone effects on laboratory indices of
  liver health: a randomized trial.
\newblock \emph{Drug and alcohol dependence}, 128\penalty0 (1-2):\penalty0
  71--76, 2013.

\bibitem[Shi et~al.(2022)Shi, Ke, Soukhavong, Lamb, Meng, Finley, Wang, Chen,
  Ma, Ye, Liu, and Titov]{lightgbm}
Yu~Shi, Guolin Ke, Damien Soukhavong, James Lamb, Qi~Meng, Thomas Finley,
  Taifeng Wang, Wei Chen, Weidong Ma, Qiwei Ye, Tie-Yan Liu, and Nikita Titov.
\newblock \emph{lightgbm: Light Gradient Boosting Machine}, 2022.
\newblock URL \url{https://CRAN.R-project.org/package=lightgbm}.
\newblock R package version 3.3.2.

\bibitem[Tibshirani(1996)]{tibshirani1996regression}
Robert Tibshirani.
\newblock Regression shrinkage and selection via the lasso.
\newblock \emph{Journal of the Royal Statistical Society: Series B
  (Methodological)}, 58\penalty0 (1):\penalty0 267--288, 1996.

\bibitem[{van der Laan}(2006)]{van2006statistical}
Mark~J {van der Laan}.
\newblock Statistical inference for variable importance.
\newblock \emph{The International Journal of Biostatistics}, 2\penalty0 (1),
  2006.

\bibitem[{van der Laan} and Gruber(2012)]{tmleLong}
Mark~J {van der Laan} and Susan Gruber.
\newblock Targeted minimum loss based estimation of causal effects of multiple
  time point interventions.
\newblock \emph{The International Journal of Biostatistics}, 8, 2012.
\newblock \doi{10.1515/1557-4679.1370}.

\bibitem[{van der Laan} and Luedtke(2015)]{van2015targeted}
Mark~J {van der Laan} and Alexander~R Luedtke.
\newblock Targeted learning of the mean outcome under an optimal dynamic
  treatment rule.
\newblock \emph{Journal of Causal Inference}, 3\penalty0 (1):\penalty0 61--95,
  2015.

\bibitem[{van der Laan} et~al.(2007){van der Laan}, Polley, and
  Hubbard]{van2007super}
Mark~J {van der Laan}, Eric~C Polley, and Alan~E Hubbard.
\newblock Super learner.
\newblock \emph{Statistical Applications in Genetics and Molecular Biology},
  6\penalty0 (1), 2007.

\bibitem[van~der Laan et~al.(2011)van~der Laan, Rose, Zheng, and van~der
  Laan]{van2011cross}
Mark~J van~der Laan, Sherri Rose, Wenjing Zheng, and Mark~J van~der Laan.
\newblock Cross-validated targeted minimum-loss-based estimation.
\newblock \emph{Targeted learning: causal inference for observational and
  experimental data}, pages 459--474, 2011.

\bibitem[{van der Laan} et~al.(2023){van der Laan}, Coyle, Hejazi, Malenica,
  Phillips, and Hubbard]{tlverseHandbook}
Mark~J {van der Laan}, Jeremy Coyle, Nima Hejazi, Ivana Malenica, Rachael
  Phillips, and Alan Hubbard.
\newblock \emph{Targeted Learning in {R}: Causal Data Science with the
  {tlverse} Software Ecosystem}.
\newblock 2023.
\newblock URL \url{https://tlverse.org/tlverse-handbook/}.

\bibitem[Wang et~al.(2016)Wang, Wu, Liu, Weng, and Zeng]{WangEHR}
Yuanjia Wang, Peng Wu, Ying Liu, Chunhua Weng, and Donglin Zeng.
\newblock Learning optimal individualized treatment rules from electronic
  health record data.
\newblock In \emph{2016 IEEE International Conference on Healthcare Informatics
  (ICHI)}, pages 65--71, 2016.
\newblock \doi{10.1109/ICHI.2016.13}.

\bibitem[Watkins(1989)]{watkins1989learning}
Christopher John Cornish~Hellaby Watkins.
\newblock Learning from delayed rewards.
\newblock 1989.

\bibitem[Weiss et~al.(2010)Weiss, Potter, Provost, Huang, Jacobs, Hasson,
  Lindblad, Connery, Prather, and Ling]{weiss2010multi}
Roger~D Weiss, Jennifer~Sharpe Potter, Scott~E Provost, Zhen Huang, Petra
  Jacobs, Albert Hasson, Robert Lindblad, Hilary~Smith Connery, Kristi Prather,
  and Walter Ling.
\newblock A multi-site, two-phase, prescription opioid addiction treatment
  study ({POATS}): rationale, design, and methodology.
\newblock \emph{Contemporary Clinical Trials}, 31\penalty0 (2):\penalty0
  189--199, 2010.

\bibitem[Weiss et~al.(2011)Weiss, Potter, Fiellin, Byrne, Connery, Dickinson,
  Gardin, Griffin, Gourevitch, Haller, et~al.]{weiss2011adjunctive}
Roger~D Weiss, Jennifer~Sharpe Potter, David~A Fiellin, Marilyn Byrne, Hilary~S
  Connery, William Dickinson, John Gardin, Margaret~L Griffin, Marc~N
  Gourevitch, Deborah~L Haller, et~al.
\newblock Adjunctive counseling during brief and extended
  buprenorphine-naloxone treatment for prescription opioid dependence: a
  2-phase randomized controlled trial.
\newblock \emph{Archives of general psychiatry}, 68\penalty0 (12):\penalty0
  1238--1246, 2011.

\bibitem[Williams and Díaz(2023)]{lmtpR}
Nicholas~T Williams and Iván Díaz.
\newblock lmtp: An {R} package for estimating the causal effects of modified
  treatment policies.
\newblock \emph{Observational Studies}, 2023.
\newblock URL \url{https://muse.jhu.edu/article/883479}.

\bibitem[Zhang et~al.(2013)Zhang, Tsiatis, Laber, and
  Davidian]{zhangSequential2013}
Baqun Zhang, Anastasios~A. Tsiatis, Eric~B. Laber, and Marie Davidian.
\newblock {Robust estimation of optimal dynamic treatment regimes for
  sequential treatment decisions}.
\newblock \emph{Biometrika}, 100\penalty0 (3):\penalty0 681--694, 2013.
\newblock \doi{10.1093/biomet/ast014}.

\bibitem[Zhao et~al.(2015)Zhao, Zeng, Laber, and Kosorok]{zhaoBowl}
Ying-Qi Zhao, Donglin Zeng, Eric~B. Laber, and Michael~R. Kosorok.
\newblock New statistical learning methods for estimating optimal dynamic
  treatment regimes.
\newblock \emph{Journal of the American Statistical Association}, 110\penalty0
  (510):\penalty0 583--598, 2015.
\newblock \doi{10.1080/01621459.2014.937488}.

\bibitem[Zhao et~al.(2012)Zhao, Zeng, Rush, and Kosorok]{owl2012}
Yingqi Zhao, Donglin Zeng, A.~John Rush, and Michael~R. Kosorok.
\newblock Estimating individualized treatment rules using outcome weighted
  learning.
\newblock \emph{Journal of the American Statistical Association}, 107\penalty0
  (499):\penalty0 1106--1118, 2012.
\newblock \doi{10.1080/01621459.2012.695674}.

\bibitem[Zhou et~al.(2017)Zhou, Mayer-Hamblett, Khan, and Kosorok]{ZhouRWL}
Xin Zhou, Nicole Mayer-Hamblett, Umer Khan, and Michael~R. Kosorok.
\newblock Residual weighted learning for estimating individualized treatment
  rules.
\newblock \emph{Journal of the American Statistical Association}, 112\penalty0
  (517):\penalty0 169--187, 2017.
\newblock \doi{10.1080/01621459.2015.1093947}.

\end{thebibliography}

\appendix
\section{Simulation}

We considered the following DGM: 

{\footnotesize
  \begin{align*}
    W_1 &\sim \mathcal{U}(-1, 1) \\
    W_2 &\sim \mathcal{U}(-1, 1) \\
    P(A_1 = 1) &= \text{expit}(0.5 - 1.3W_1 + 0.4W_2) \\
    W_3 &\sim \mathcal{U}(-1, 1) \times 1.25A_1 + 0.25\\
    P(A_2 = 1) &= \text{expit}(0.5 + 0.4A_1 - 1.5W_3) \\
    Y &\sim \mathcal{N}(0.4 - 0.4A_1 - A_2W_3 - 4A_1W_1 + 0.08A_1A_2 + A_2W_3 - 4A_1W_1 - 2A_1W_2 - 0.1A_2 + 1.5W_1, 1) \\
  \end{align*}
 }

We evaluated estimator performance in terms of absolute bias, absolute bias scaled by $\sqrt{n}$, and 95\% confidence interval (CI) coverage. We conducted 1000 simulations for sample sizes $n \in \{500, 1000, 10000\}$. The true value of $\psi$ is approximately $0.248$.

\begin{table}[H]
\centering
\footnotesize
\caption{Simulation results.}
\begin{tabular}[t]{lccccc}
\toprule
$n$ & $\hat{\psi}$ & $|\text{Bias}|$ & $\sqrt{n} \times |\text{Bias}|$ & 95\% CI Covr. \\
\midrule
500 & 2.269 & 0.023   & 0.514 & 0.929   \\
1000 & 2.267 & 0.021  & 0.653 & 0.925  \\
10000 & 2.250 & 0.004 & 0.378 & 0.875 \\
\bottomrule
\end{tabular}
\end{table}

\end{document}